\documentclass[aps,prb,twocolumn,amsmath,amssymb,superscriptaddress,floatfix]{revtex4-1}
\usepackage{amsmath}
\usepackage{amssymb}
\usepackage{amsfonts}
\usepackage{listings}
\usepackage{enumerate}
\usepackage{latexsym}
\usepackage{bm}
\usepackage{graphicx}
\usepackage{braket}
\usepackage[pdftex,colorlinks=false]{hyperref}
\usepackage{xcolor}
\usepackage{verbatim}

\usepackage{times}

\newcommand{\beq}{\begin{equation}}
\newcommand{\eneq}{\end{equation}}

\def\be{\begin{equation}}
\def\ee{\end{equation}}
\def\ba{\begin{eqnarray}}
\def\ea{\end{eqnarray}}

\def\R{{\rm Re}}
\def\Z{\mathbb{Z}}
\def\C{\mathbb{C}}

\def\Tr{{\rm Tr}}

\newcommand{\eqnref}[1]{Eq.~\eqref{#1}}

\def\beq{\begin{equation}}
\def\eeq{\end{equation}}
\def\barray{\begin{eqnarray}}
\def\earray{\end{eqnarray}}

\font\upright=cmu10 scaled\magstep1
\def\stroke{\vrule height8pt width0.4pt depth-0.1pt}

\def\Zmath{\mathbb{Z}}
\def\Qmath{\vcenter{\hbox{\upright\rlap{\rlap{Q}\kern
                   3.8pt\stroke}\phantom{Q}}}}
\def\Nmath{\vcenter{\hbox{\upright\rlap{I}\kern 1.7pt N}}}
\def\Cmath{\vcenter{\hbox{\upright\rlap{\rlap{C}\kern
                   3.8pt\stroke}\phantom{C}}}}
\def\Rmath{\vcenter{\hbox{\upright\rlap{I}\kern 1.7pt R}}}
\def\Z{\ifmmode\Zmath\else$\Zmath$\fi}
\def\Q{\ifmmode\Qmath\else$\Qmath$\fi}
\def\N{\ifmmode\Nmath\else$\Nmath$\fi}
\def\C{\ifmmode\Cmath\else$\Cmath$\fi}
\def\R{\ifmmode\Rmath\else$\Rmath$\fi}

\input{epsf}

\newcounter{defcounter}
\begin{document}

\tolerance 10000

\newcommand{\cbl}[1]{\color{blue} #1 \color{black}}

\newcommand{\vk}{{\bf k}}

\widowpenalty10000
\clubpenalty10000

\title{Measurement of the entanglement spectrum of a symmetry-protected topological state using the IBM quantum computer}

\author{
Kenny~Choo}
\address{
 Department of Physics, University of Zurich, Winterthurerstrasse 190, 8057 Zurich, Switzerland
}

\author{
Curt~W.~von~Keyserlingk}
\address{
University of Birmingham, School of Physics and Astronomy, B15 2TT, UK
}

\author{Nicolas Regnault}
\address{
Laboratoire Pierre Aigrain, Ecole normale sup\'erieure, PSL University, 
Sorbonne Universit\'e, Universit\'e Paris Diderot, Sorbonne Paris Cit\'e, 
CNRS, 24 rue Lhomond, 75005 Paris France
}

\author{
Titus~Neupert}
\address{
 Department of Physics, University of Zurich, Winterthurerstrasse 190, 8057 Zurich, Switzerland
}

\begin{abstract}
Entanglement properties are routinely used to characterize phases of quantum matter in theoretical computations.
For example the spectrum of the reduced density matrix, or so-called "entanglement spectrum", has become a widely used diagnostic for universal topological properties of quantum phases.  
However, while being convenient to calculate theoretically, it is notoriously hard to measure in experiments.
Here we use the IBM quantum computer to make the first ever measurement of the entanglement spectrum of a symmetry-protected topological state. We are able to distinguish its entanglement spectrum from those we measure for trivial and long-range ordered states.
\end{abstract}

\date{\today}

\maketitle

\textit{Introduction ---}
The patterns of entanglement between local degrees of freedom are a fingerprint of many-body quantum phases \cite{ChenChengWen2013, KitaevPreskill2006,WenZeng15}. Various measures for quantum entanglement capture different universal aspects of quantum states and allow for their classification. Of these measures, the entanglement spectrum, obtained from the spectrum of the reduced density matrix of a bipartitioned quantum system, reveals the most detailed information. It was first used in this way by Li and Haldane \cite{LiHaldane2008}, who characterized the topological order of a fractional quantum Hall system by matching its low-lying entanglement level counting to the universal level structures of conformal field theories. 

Entanglement spectroscopy further became an important tool in identifying symmetry protected topological states of matter (SPTs) \cite{ChenChengWen2013, KitaevPreskill2006, LevinWen2006, PollmannTurner2010, ChenGuWen2011a, ChenGuWen2011b}.  The defining property of an SPT is that it cannot be connected to trivial product state via a finite-depth quantum circuit of local symmetry preserving unitary operations. SPTs are not topologically ordered in the sense of the fractional quantum Hall effect, but still feature a topological bulk-boundary correspondence: (in one dimension) they always support gapless boundary excitations in an open geometry, as long as the boundary does not break the protecting symmetry. This bulk boundary correspondence is also manifested in the entanglement spectrum: the bipartitioning used to define the reduced density matrix introduces a `virtual' boundary between parts A and B of the system, and in an SPT topological excitations are associated with this virtual boundary leading to protected degeneracies in the entanglement spectrum \cite{Fidkowski2010}.

While entanglement spectra conveniently characterize many-body quantum states in numerical simulations, they are intrinsically hard to measure in condensed matter experiments. It is only recently with the improvements in trapped ions and superconducting quantum simulators that the entanglement entropy~\cite{GreinerEntanglementEntropy2015} or density matrix ~\cite{ BlattLanyon2018} of many-spin systems can be accessed~\cite{JohriSteigerTroyer2017}. The recent release of IBM cloud quantum computing service allows to test various conceptual ideas on an actually existing quantum simulator~\cite{IbmQuantumExperience}. 
Several recent works characterize the entanglement properties of the IBM devices~\cite{WangLiYin2018}, use them to test various quantum algorithms~\cite{Alsina2016,Devitt2016,Rundle2017, Huffman2017,Hebenstreit2017}, and to solve physical demonstration problems~\cite{KandalaIBMMolecule2017,ArrowOfTime2017}.
Here we show that the high level of control that can be reached on the IBM analogue quantum computer allows one to both prepare an SPT state, and to measure its entanglement spectrum.


\begin{figure*}[t]
            \includegraphics[width=0.9\textwidth]{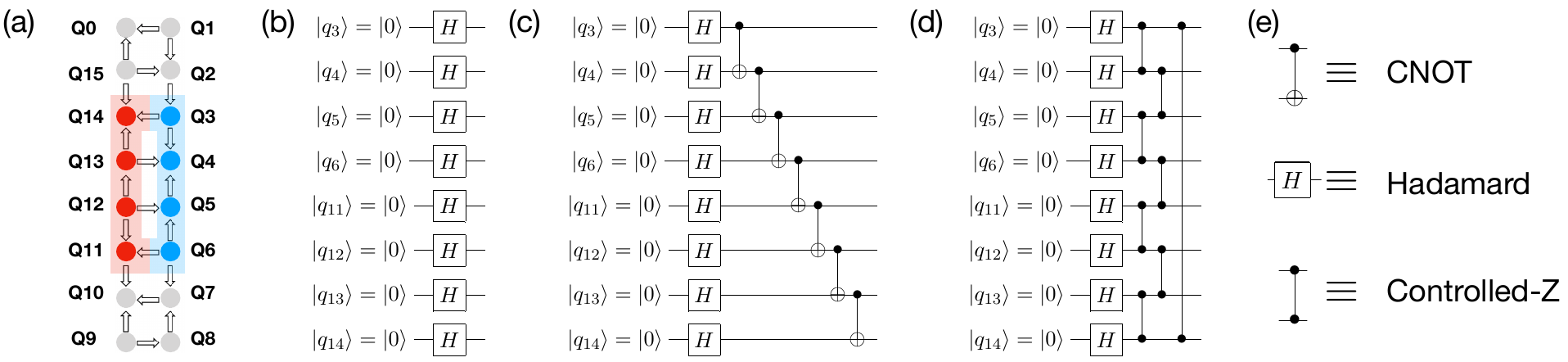}
    \caption{
    (a) Layout of the qubits and two-qubit gates (arrows) in IBM quantum computer ibmqx5. 
    Blue and red qubits were used to measure entanglement spectra and formed subsystem $A$ and $B$ of the simulated quantum spin chain (with periodic boundary conditions), respectively. 
    (b)--(d) Circuits used for constructing the quantum states, where H stands for a Hadamard and $+$ for a CNOT gate, respectively. 
    (b) Constructs the ground state of a trivial paramagnet.
    (c) Constructs the eight qubit cat state (also `GHZ state').
    (d) Constructs the ground state of a topological paramagnet (also `graph state'). Hadamard gates that appear next to each other are automatically removed.
    (e) Symbols for the various logic gates.
    }
\label{fig: circuits}
\end{figure*}

\textit{Topology in quantum paramagnets ---}
SPT phases are gapped short ranged entangled states with a protecting symmetry. 
In this paper, the we are interested in SPT phases of one-dimensional paramagnets arising in a chain of spin-1/2 degrees of freedom with local quantum states $\mid\uparrow\rangle$ and $\mid \downarrow\rangle$ at each site (corresponding to a chain of qubits on the IBM quantum computer). 
According to the classification of SPT phases~\cite{ChenChengWen2013}, such one-dimensional bosonic systems 
support a topologically nontrivial phase protected by time-reversal symmetry $\mathcal{T}$ with $\mathcal{T}^2=+1$. 
We choose the representation $\mathcal{T}=K\prod_{i} \sigma_{x,i}$, where $K$ is complex conjugation and $\sigma_{\mu,i}$ is the Pauli operator with $\mu=0,x,y,z$ acting on site $i$ of the chain ($\sigma_0$ being the identity operator). 

A trivial paramagnetic state invariant under $\mathcal{T}$ is given by the ground state of
\begin{equation}
H_{\mathrm{triv}} = - \sum_{i} \sigma_{x,i}
\end{equation}
which for the example of a chain of length $N$, is a trivial product state
\begin{equation}
\ket{\mathrm{PM}} = \frac{1}{\sqrt{2^{N}}} \sum_{\boldsymbol{r}} \ket{\boldsymbol{r}} = \ket{+}^{\otimes ^N},
\label{eq: ES definition}
\end{equation}
where $\boldsymbol{r}$ represents all possible binary strings of length $N$ and $\ket{+}$ is the eigenstate of $\sigma_{x}$ with eigenvalue $+1$. 

On the other hand, a topologically nontrivial SPT state protected by $\mathcal{T}$ is given by the ground state of the stabilizer Hamiltonian 
\begin{equation}
H_{\mathrm{topo}} = -\sum_{i} \sigma_{z,i-1}\sigma_{x,i}\sigma_{i+1}
\label{eq: SPT Hamiltonian}
\end{equation}
with periodic boundary conditions imposed. Since all terms in $H_{\mathrm{topo}}$ commute, its unique ground state $\ket{\mathrm{SPT}}$ is defined by 
\begin{equation}
\sigma_{z,i-1}\sigma_{x,i}\sigma_{z,i+1}\ket{\mathrm{SPT}}=\ket{\mathrm{SPT}},\qquad \forall i.
\label{eq: psi definition}
\end{equation}
This state is the so called graph state which can be use as a key resource to achieve one-way quantum computation \cite{RaussendorfBriegel2001,BriegelRaussendorf2001}. It is also topologically equivalent to the well-known Affleck-Kennedy-Lieb-Tasaki (AKLT) state \cite{AKLT1987,AKLT1988}.

We now discuss how to distinguish the trivial \eqnref{eq: ES definition} and non-trivial SPT ground states \eqnref{eq: psi definition}. The classification of one-dimensional SPT phases in Ref.~\onlinecite{ChenChengWen2013} is based on the fact that in one dimension any paramagnetic/short-range correlated state can be transformed into a trivial product state via a local unitary (LU) transformation. However, if the paramagnetic state is a non-trivial SPT, some of the local operations comprising such a LU transformation necessarily break the protecting symmetry. LU transformations are well approximated by finite depth quantum circuits \cite{Chen2010}. To define a quantum circuit, we first introduce the piecewise local unitary operators 
\begin{equation}
\tilde{U}_{\mathcal{P}} = \prod_{i\in \mathcal{P}}  U_{i}
\label{eq: piecewise local U}
\end{equation}
where ${U_{i}}$ is a set of unitary operations acting on \emph{nonoverlapping} local regions listed in the set $\mathcal{P}$. For the IBM quantum computer the only available multi-qubit gate is a CNOT gate, that is, $U_{i}$ acts on sites $i$ and $i+1$.
 The full LU transformation is then given by a finite product of $M$ piecewise unitaries of the form of Eq.~\eqref{eq: piecewise local U}
\begin{equation}
U_{\mathrm{LU}}^{M} = \tilde{U}_{\mathcal{P}_1} \cdots \tilde{U}_{\mathcal{P}_M}.
\end{equation}

\begin{figure*}[t]
\begin{center}
            \includegraphics[width=0.98\textwidth ]{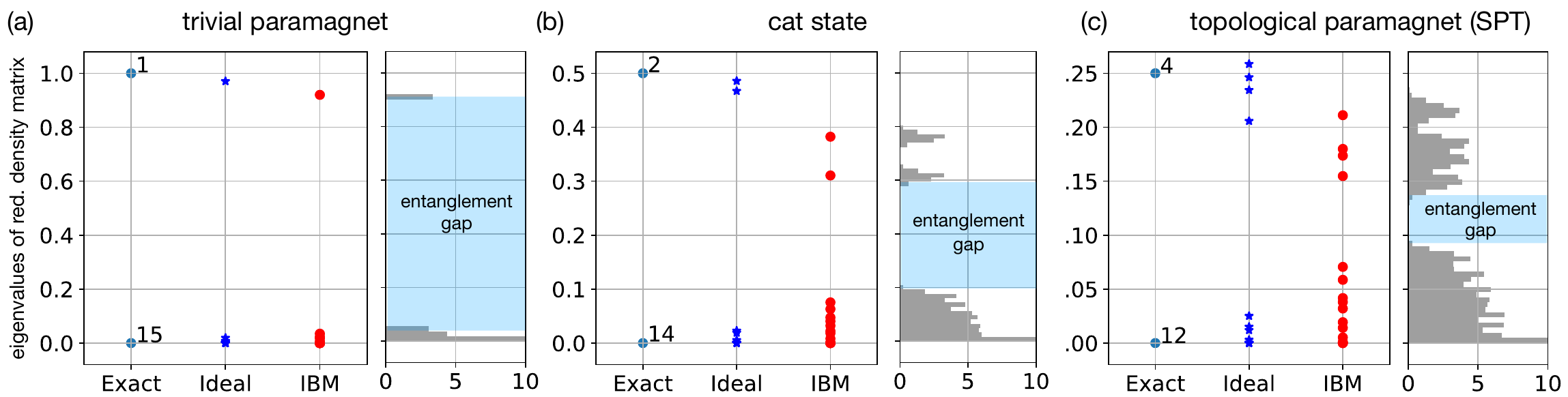}
     
    \caption{Entanglement spectra measured on the IBM quantum computer (red) compared to the theoretical expectations. Light blue symbols are the theoretical expectations for the respective ideal quantum state. The blue symbols 
   include the effects of sampling noise, present even for an ideal quantum state, ideal gates and no readout error. Both for the simulated and for the actual measurement, each observable was measured 1024 times. The plots show the eigenvalues of the reduced density matrix, the largest eigenvalues correspond to the lowest levels in the entanglement spectrum.    
    (a) Ground state of a trivial paramagnet: A unique lowest state is found in the entanglement spectrum, reflecting the absence of any entanglement between regions A and B.
    (b) Eight qubit cat state: A doubly degenerate lowest entanglement level is found reflecting the long range correlations in this state.
    (c) Ground state of a topological paramagnet: An entanglement gap separating four low-lying levels is the fingerprint of this topologically nontrivial state. The theoretically expected perfect degeneracy is already substantially lifted by the statistical noise (blue), and in the actual experiment further compromised by decoherence, gate errors and readout errors.
    The right subpanels are eigenvalues of reduced density matrices drawn from the statistical distribution of them when accounting for the statistical noise associated with the measurement.
    }
\label{fig: entanglement spectra}
\end{center}
\end{figure*}

A LU transformation that transforms the SPT state $\ket{\mathrm{SPT}}$ [Eq.~\eqref{eq: psi definition}] into the trivial product state $\ket{\mathrm{PM}}$  [Eq.~\eqref{eq: ES definition}] is obtained by choosing 
\begin{equation}
U_{i}=\frac{1}{2}\sigma_{0,i}(\sigma_{0,i+1}+\sigma_{z,i+1})+\frac{1}{2}\sigma_{z,i}(\sigma_{0,i+1}-\sigma_{z,i+1}),
\end{equation}
which represents a controlled-Z gate between spin $i$ and $i+1$. Further using $M=2$ with $\mathcal{P}_1$ and $\mathcal{P}_2$ containing all even and all odd sites, respectively, one obtains
\begin{equation}
\ket{\mathrm{PM}} = \prod_{i}  U_{i}  \ket{\mathrm{SPT}}
\end{equation}
as the LU transformation. Since $U_{i}^2=\openone$, the same LU transformation can be used to construct the state $ \ket{\mathrm{SPT}}$ from $\ket{\mathrm{PM}}$.
The  quantum circuit corresponding to $\prod_{i}  U_{i}$ is shown in Fig.~\ref{fig: circuits}(d). It is straightforward to see that none of the $U_{i}$ commute with  $\mathcal{T}$, as expected for a nontrivial SPT state $\ket{\mathrm{SPT}}$ and a trivial product state $\ket{\mathrm{PM}}$.

Another characterization of the SPT state is that with open boundary conditions Hamiltonian~\eqref{eq: SPT Hamiltonian}  supports gapless excitations localized at the chain end. To see this, notice that the operators $\sigma_{x,1}\sigma_{z,2}$, $\sigma_{y,1}\sigma_{z,2}$, and $\sigma_{z,1}$ all commute with the Hamiltonian and form a Pauli algebra. They thus enforce a two-fold degeneracy of the ground state associated with a localized excitation at the boundary. Since all these operators are odd under $\mathcal{T}$ they cannot be added as perturbations to the Hamiltonian and we conclude that the gapless topological end excitation is protected by $\mathcal{T}$.

\textit{Entanglement spectrum ---}
For any quantum state $\ket{\psi}$, the reduced density matrix of subsystem $A$ (some subset of the lattice ) is given by the partial trace of the full density matrix $\rho = \ket{\psi}\bra{\psi}$ over the complementary subsystem $B=A^{c}$, i.e.,
\begin{equation}
\rho_{A} = \Tr_{B} \rho.
\end{equation}
The entanglement spectrum is then the set of numbers $\lambda_i$, $i=1,\cdots,N_{A}$, where $e^{-\lambda_i}$ are the eigenvalues of the operator $\rho_{A}$ and $N_{A}$ is the dimension of the Hilbert space describing all degrees of freedom in subsystem $A$.
The entanglement spectrum can reveal information about universal, in particular topological, properties of the state $\ket{\psi}$. For example, if the bipartitioning is realized by a cut  separating a linear chain into a left and a right half, it has been argued \cite{PollmannTurner2010} that nontrivial SPT phase have entanglement spectra whose low energy spectra are in one-to-one correspondence with the spectrum of the system in the presence of a boundary. This statement can be proved straightforwardly for non-interacting free fermion systems \cite{Fidkowski10}.
For the case at hand, we study a chain with periodic boundary boundary conditions, so that a an entanglement cut introduces two `virtual' boundaries between parts $A$ and $B$. If each cut comes with a twofold degeneracy, we expect a $2\times2=4$-fold degenerate entanglement spectrum of the SPT. In contrast, no degeneracies are expected in the low-lying entanglement spectrum of the trivial PM phase, as it has no protected boundary modes.
The degeneracy is thus a useful diagnostic for identifying topological phases.

We obtain the entanglement spectrum by measuring the reduced density matrix of the subsystem. Given any reduced density matrix $\rho_{A}$ we can decompose it into a sum of Pauli matrices. For a $n$-spin subsystem at sites $1,\cdots,n$ we have
\begin{equation} \label{redDM}
\rho_{A} = \sum_{\alpha_1\cdots\alpha_n} c_{\alpha_1\cdots\alpha_n} \frac{1}{2^n}\sigma_{\alpha_1,1} \sigma_{\alpha_2,2}  \cdots \sigma_{\alpha_n,n}
\end{equation}
where the $\frac{1}{2^n}$ is a normalization factor and the indices $\alpha_i, \ i=1,\cdots, n$, run over the set $\lbrace 0, x, y, z\rbrace$. The aim then is to obtain the coefficients $c_{\alpha_1\cdots\alpha_n}$ given that we have access to the full density matrix. Since the Pauli matrices are orthogonal under the trace norm, the coefficients are given by
\begin{equation}
\begin{split}
c_{\alpha_1\cdots\alpha_n} 
&= \Tr_{A}\left[\sigma_{\alpha_1,1} \sigma_{\alpha_2,2} \cdots  \sigma_{\alpha_n,n} \rho_{A}\right] \\
&= \Tr_{A}\left[\sigma_{\alpha_1,1} \sigma_{\alpha_2,2}  \cdots \sigma_{\alpha_n,n} \Tr_{B}(\rho)\right] \\
&= \Tr_{A} \Tr_{B} \left[\sigma_{\alpha_1,1} \sigma_{\alpha_2,2}  \cdots  \sigma_{\alpha_n,n} \rho \right] \\
&= \Tr \left[\sigma_{\alpha_1,1} \sigma_{\alpha_2,2} \cdots \sigma_{\alpha_n,n} \rho \right].
\end{split}
\end{equation}
This means that in order to obtain the coefficients $c_{\alpha_1\cdots\alpha_n}$, we need only measure all the Pauli operators corresponding to subsystem $A$. For 4 spins, and ignoring the identity operator, this gives $4^{4}-1 = 255$ measurements.

After obtaining the coefficients, we reconstruct the reduced density matrix according to Eq.~\eqref{redDM} and we numerically diagonalize $\rho_{A}$ to obtain the entanglement spectrum.

\textit{Results---}
We consider the different quantum states on a system of 8 spins (qubits):
The trivial paramagnet $\ket{\textrm{PM}}$ from Eq.~\eqref{eq: ES definition}, the 
cat state
\begin{equation}
\ket{\textrm{Cat}} = \frac{1}{\sqrt{2}} (\ket{00000000} + \ket{11111111})
\end{equation}
and the SPT state (graph state) as defined in Eq.~\eqref{eq: psi definition}.
We construct these states on the IBM 16 qubit quantum computer \emph{ibmqx5} using the quantum circuits shown in Fig.~\ref{fig: circuits}.

To obtain the reduced density matrix, we have to obtain the expectation values of the Pauli operators according to Eq.~\eqref{redDM}. Since each measurement gives only a $0$ or $1$, the experiment  consisting of state preparation and measurement is repeated multiple times in order to estimate the expectation value. The results we show are taken with 1024 repetitions all executed within about 15 minutes on the quantum computer.  We refer to the supplementary information for performance characteristics of the quantum computer (as well as further information about the measurements. 

There are several sources of error which prevent us from exactly reproducing the density matrix. These errors could lead to unphysical density matrices with negative eigenvalues. The standard way to avoid this is to use a maximum likelihood estimator \cite{SmolinGambetta2012} to obtain the closest physical density matrix based on the estimated one. All entanglement spectra obtained is shown in Fig.~\ref{fig: entanglement spectra} (In the supplementary material, we provide the calibration data of the ibmqx5 for the day where the measurement of Fig. 2c was taken. We also show in Fig.~\ref{fig: graph samples} the entanglement spectra obtained another day and thus with another calibration set. While there is some variation between measurements at different days, the qualitative picture remains unchanged.). We now list the various sources of error. 

The repeated measurement gives rise to a statistical noise. For example, if the observable $\sigma_{z}$ in the state $\ket{+} = \frac{1}{\sqrt{2}} (\ket{0} + \ket{1})$ is measured 100 times, within one standard deviation, one may obtain 55 times '0' and 45 times '1', and would conclude that $\langle\sigma_{z}\rangle=0.1$ instead of the exact 0. To estimate the scale of the latter error, we computed the probabilistic outcome of $1024\times256$ measurements on the \emph{exact} states for each of the three cases $\ket{\mathrm{PM}}$, $\ket{\mathrm{Cat}}$, $\ket{\mathrm{SPT}}$. The entanglement spectra computed from the resulting estimate of the density matrix are plotted in Fig.~\ref{fig: entanglement spectra} in the column `Ideal'.  

\begin{figure}[t]
\begin{center}
            \includegraphics[width=0.40\textwidth ]{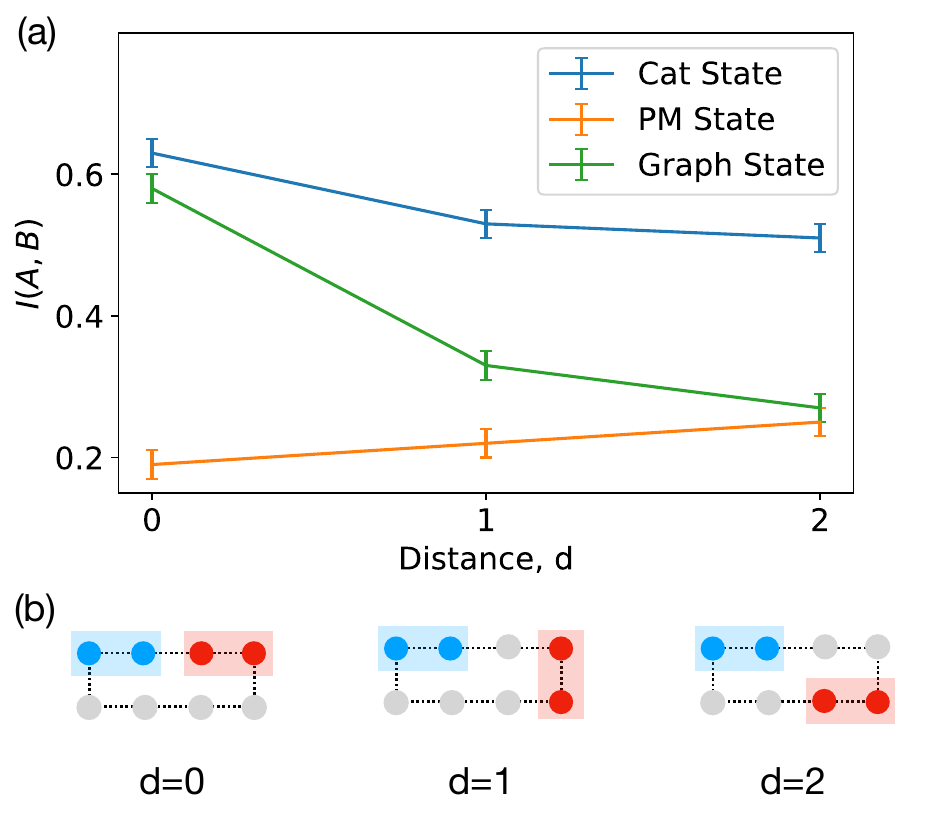}
    \caption{(a)Measured mutual information between two regions of the 8-site spin chain.
    The results are obtained from the measured reduced density matrices, measured in the same way as for Fig.~\ref{fig: entanglement spectra}.
    The theoretically exact values are $I(A,B)=0$ and $I(A,B)=\mathrm{log}2$ for all $d$ in the case of $|\mathrm{PM}\rangle$ and $|\mathrm{Cat}\rangle$, respectively, and $I(A,B)=2\mathrm{log}2,\mathrm{log}2,0$ for $|\mathrm{SPT}\rangle$ at distances $d=0,1,2$, respectively.  (b) Schematic depiction of Subsystem A (Blue) and Subsystem B (red) for the various distances $d=0,1,2$.
    }
\label{fig: mutual information}
\end{center}
\end{figure}

Additionally, errors in the real quantum computer arise from a finite coherence time of the qubits, as well as errors involved in implementing the gates and in reading out the qubits. They account for the difference between the columns `Ideal' and `IBM' (the latter obtained from the actual measurement) in Fig.~\ref{fig: entanglement spectra}. To quantify the statistical noise in the latter, we have generated a distribution of eigenvalues of reduced density matrices drawn from their distribution (see supplemental material for details). The distributions are shown in the respective right subpanels in Fig.~\ref{fig: entanglement spectra}. They show a robust entanglement gap in each of the cases $\ket{\mathrm{PM}}$, $\ket{\mathrm{Cat}}$, $\ket{\mathrm{SPT}}$, with the expected one, two, and four states above the gap. The four-fold degeneracy of the entanglement spectrum of $\ket{\mathrm{SPT}}$ is only approximate in the measurement due to a combination of statistical noise and more importantly also the errors present in the quantum computer. 

Using the measured reduced density matrices, we can further compute various entanglement measures. In the supplemental Sec.~\ref{supp: entanglement}, we give the von Neumann entropy $S_{1,A}=-\mathrm{Tr}_A[\rho_A\,\log \rho_A]$ and the second Renyi entropy $S_{2,A}=-\log\,\mathrm{Tr}_A\,[\rho_A^2]$. The entanglement structure of gapped ground states of local Hamiltonians in one dimensional systems is short-ranged. This property can be studied via the mutual information
\begin{equation}
I(A,B)=S_{1,A}+S_{1,B}-S_{1,AB},
\end{equation}
where $A$ and $B$ are two subsystems. For the 8-site chain that we study, $A,B$ consist of two adjacent sites each. $A,B$ can then be a distance $d=0$, $d=1$ or $d=2$ sites apart, owing to the periodic boundary conditions. We measured the density matrix on the four sites comprising subsystems $A,B$ for $d=0,1,2$ in each of the three prepared states $|\mathrm{PM}\rangle$, $|\mathrm{Cat}\rangle$ and $|\mathrm{SPT}\rangle$. The results, summarized in Fig.~\ref{fig: mutual information}, are qualitatively consistent with the following theoretical facts: (i) The mutual information of $|\mathrm{Cat}\rangle$ cannot be a gapped ground state of a local Hamiltonian, as its mutual information does not decay with distance, and (ii) $|\mathrm{SPT}\rangle$ has only local entanglement that decays with distance. Furthermore, (iii)  $|\mathrm{PM}\rangle$ is a local product state with vanishing correlations, as even for $d=0$ the mutual information vanishes; in our data, this mutual information does not vanish, but it is certainly smaller compared to the two other states. 

\textit{Summary---}
We measured the entanglement spectrum, a universal fingerprint of topological phases of matter, using the IBM quantum computer. 
With the quickly increasing gate fidelities and qubit coherence times \cite{Barends:2016aa, Kelly:2015aa}, it seems only a matter of time before quantum simulators/quantum computers will be able to make precise measurements of the entanglement spectrum of larger systems, thus providing a useful tool in understanding topological phases of matter.
Further, since `topological software', for instance the surface (toric) code~\cite{KitaevToric2003}, is envisioned as one venue to turn an analogue quantum computer into a digital (error-corrected) quantum computer, the characterization of quantum states in terms of their entanglement spectrum will be indispensable.

\textit{Acknowledgments ---}
KC was supported by the European UnionÕs Horizon 2020 research and innovation program (ERC-StG-Neupert-757867-PARATOP).
CWK was supported by a Birmingham Fellowship.

\bibliography{biblio}

\clearpage
\begin{widetext}
\section*{Supplementary Information}

\subsection{Performance of the quantum computer and additional measurements}
\label{supp: performance}

All measurements in this paper were done on the IBM 16-qubit quantum computer \textit{ibmqx5} which is calibrated twice daily. The single qubit gate errors are generally of the order $\sim 0.2\%$  while the controlled-not gates have an error $\sim4\%$. In the table below, we show the specific calibration of the device at the time when graph state density matrix in Fig.~\ref{fig: entanglement spectra} was measured.
\begin{table}[h]
\begin{center}
\setlength{\tabcolsep}{.5em}
\renewcommand{\arraystretch}{1.5}
\begin{tabular}{ c|cccccccccccccccc } 
 \hline\hline
 Qubit & $Q_{0}$ & $Q_{1}$ & $Q_{2}$ & $Q_{3}$ & $Q_{4}$ & $Q_{5}$ & $Q_{6}$ & $Q_{7}$ & $Q_{8}$ & $Q_{9}$ & $Q_{10}$ & $Q_{11}$ & $Q_{12}$ & $Q_{13}$ & $Q_{14}$ & $Q_{15}$ \\ 
 \hline
 $T_{1}$ in $\mu$s                        & 47.1 & 35.3 & 35.5 & 55.3 & 48.5 & 51.2 & 45.3 & 29.6 & 45.5 & 56.9 & 59.7 & 31.6 & 59.5 & 45.8 & 33.0 & 51.2\\ 
 $T_{2}$ in $\mu$s                        & 36.0 & 53.8 & 53.2 & 63.4 & 81.9 & 49.0 & 79.9 & 34.8 & 71.6 & 108.1 & 85.9 & 28.3 & 53.4 & 86.6 & 53 & 96.3\\ 
 $\epsilon_{r}$$\times 10^{2}$           & 8.39 & 6.31 & 4.95 & 4.54 & 8.89 & 4.73 & 5.31 & 4.83 & 4.23 & 8.83 & 7.65 & 4.68 & 11.58 & 4.83 & 5.83 & 9.22\\ 
 $\epsilon_{\sigma}$$\times10^{3}$  & 2.05 & 3.42 & 3.57 & 1.91 & 1.20 & 1.93 & 1.60 & 1.92 & 1.45 &  0.92 & 1.64 & 2.21 & 1.50 & 1.51 & 2.08 & 3.00\\ 
 \hline\hline
\end{tabular}
\end{center}
\caption{\textbf{Qubit and readout characterization.} Relaxation time $T_{1}$, coherence time $T_{2}$, readout error $\epsilon_{r}$ and single qubit gate error $\epsilon_{\sigma}$. These values correspond to the calibration of the IBM \textit{ibmqx5} 16-qubit quantum computer at the time when the graph state density matrix in Fig.~\ref{fig: entanglement spectra} was measured.}
\end{table}

The graph state density matrix in Fig. 2 was measured in a timeframe that involve a single calibration set. To avoid any bias, we have repeated such a determination at different times and thus for different calibration sets. While there is some variation between measurements at different days, the qualitative picture is unchanged as can be observed in Fig. 4. However, note that it can happen (if one uses bad qubits or if the device was not calibrated as well on a particular day) that the qubit and gate fidelities are too poor to observe a clear entanglement gap.

\begin{figure}[h]
\begin{center}
            \includegraphics[width=0.40\textwidth ]{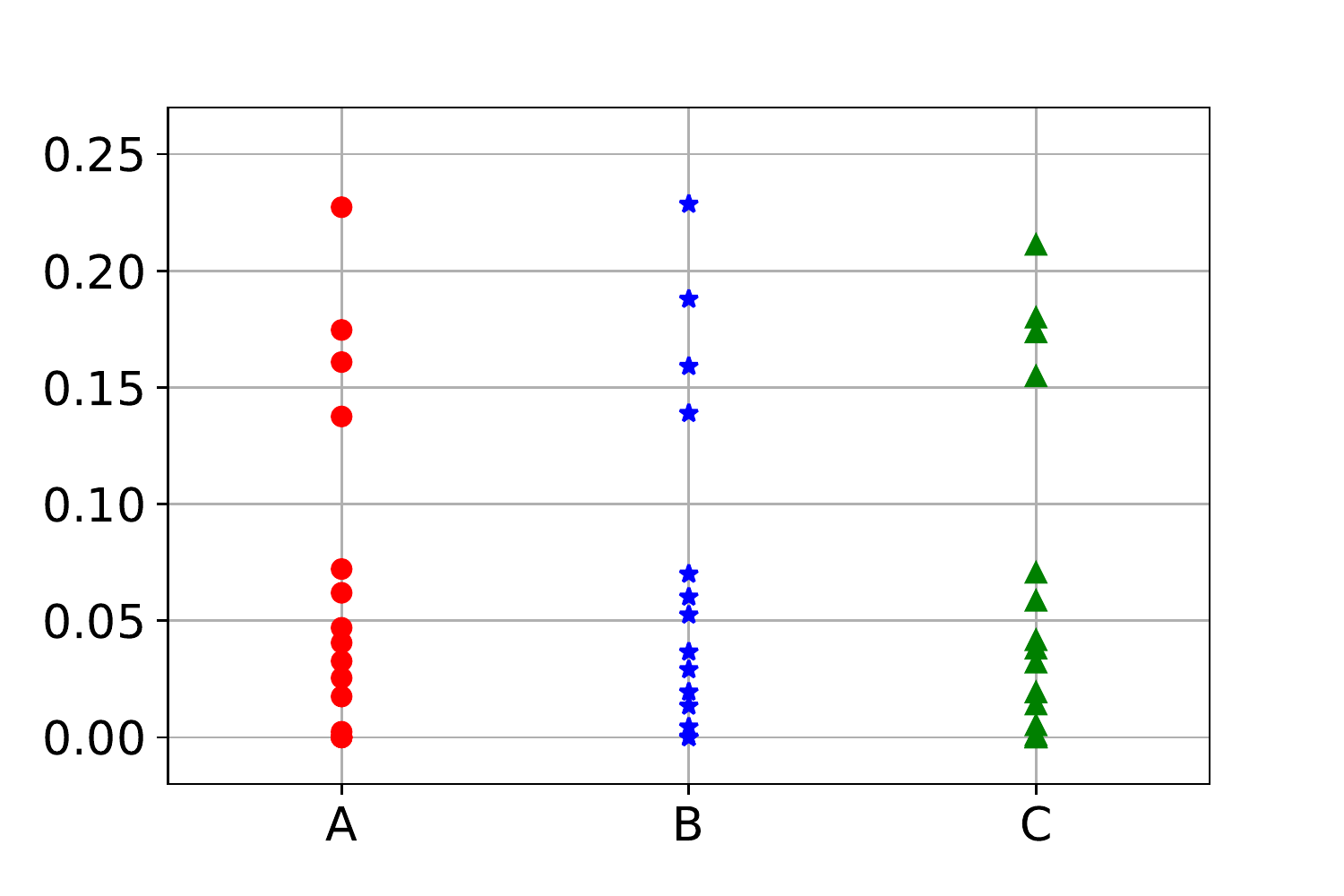}
    \caption{Some samples of the graph state entanglement spectra taken on different days. Sample C correspond to the measurement shown in the main text.
    }
\label{fig: graph samples}
\end{center}
\end{figure}

\subsection{Measurements and Error estimation}
\label{supp: error}
In order to measure the reduced density matrix of a subsystem $A$, we consider its decomposition into a sum of Pauli operators
\begin{equation} \label{redDM}
\rho_{A} = \sum_{\alpha_1\cdots\alpha_n} c_{\alpha_1\cdots\alpha_n} \frac{1}{2^n}\sigma_{\alpha_1,1} \sigma_{\alpha_2,2}  \cdots \sigma_{\alpha_n,n}
\end{equation}
the coefficients are then given by the expectation value with respect to the full system $c_{\alpha_1\cdots\alpha_n} = \textrm{Tr}(\sigma_{\alpha_1,1} \sigma_{\alpha_2,2}  \cdots \sigma_{\alpha_n,n} \rho)$. 

To estimate the coefficients $c_{\alpha_1\cdots\alpha_n}$, we prepare the state and measure the corresponding Pauli operator, the result of which would be $1$ with a probability $p$, and $-1$ with probability $(1-p)$. We perform the measurement $n$ times, where we typically use $n=1024$ shots. This gives rise to a binomial distribution $P(k)$ where $k$ is the number of $1$'s obtained. The coefficient can then be estimated by 
\begin{equation}
\hat{c}_{\alpha_1\cdots\alpha_n} = \frac{k - (n-k)}{n} = 2\frac{k}{n} -1.
\end{equation}
Since the variable $k$ is distributed according to a binomial distribution we can compute the uncertainty in the estimate $\hat{c}_{\alpha_1\cdots\alpha_n}$ as
\begin{equation}
\delta_{\alpha_1\cdots\alpha_n} = \frac{2}{n}  \sqrt{\textrm{Var}(k)}= \frac{2}{\sqrt{n}}  \sqrt{p(1-p)}.
\end{equation}

However, this uncertainty is in the coefficients $c_{\alpha_1\cdots\alpha_n}$ of the density matrix and does not easily translate into an uncertainty of the individual eigenvalues. Therefore, to illustrate the uncertainty in the eigenvalues, we sample the coefficient with a Gaussian distribution with mean $c_{\alpha_1\cdots\alpha_n}$ and standard deviation $\delta_{\alpha_1\cdots\alpha_n}$. We then construct the density matrix with this new coefficients and obtain a new eigenspectrum. Repeating this procedure many times gives a probability density distribution of eigenvalues shown in Fig.~\ref{fig: entanglement spectra} on the right side of each panel. We use the same method to compute the uncertainty in entanglement entropies.

\subsection{Projection on physical density matrices}
\label{supp: projection}
Any $d\times d$ density matrix can be decomposed into an orthonormal Hermitian operator basis $\lbrace{ \sigma_{i} \rbrace}^{d^2}_{i=1}$ [with $\textrm{Tr}(\sigma_{i}\sigma_{j})=d \delta_{ij}$]
\begin{equation} \label{redDM}
\mu = \frac{1}{d} \sum_{i} c_{i} \sigma_{i},
\end{equation}
where the $\frac{1}{d}$ is a normalization factor. The coefficients $c_{i}$ are then given by the expectation value $c_{i} = \textrm{Tr}(\sigma_{i} \mu)$. For generic coefficients this matrix would not satisfy the conditions for being a physical density matrix, namely: positive semidefinite, Hermitian and unit trace.

In order to estimate a physical density matrix from a generic unphysical one, we used maximum likelihood estimation \cite{SmolinGambetta2012}, which is implemented in the IBM QISkit library. The basic idea is as follows: Given a set of coefficients $c_{i}$ we want to find the physical density matrix $\rho$ which minimizes the negative log likelihood function using a Gaussian prior
\begin{equation} \label{eq: logL}
\mathcal{L}_{\log} = \sum_{i} \left[ c_{i} - \textrm{Tr}(\sigma_{i}\rho) \right]^{2} = \textrm{Tr}(\mu - \rho)^2 = \sum_{ij} |\mu_{ij} - \rho_{ij}|^{2}.
\end{equation}
This corresponds to the 2-norm which is basis independent, therefore we can switch to the eigenbasis of $\mu$ and the problem is transformed into a least squares minimization.
In this basis $\mu$ is diagonal. The optimal $\rho$ is also diagonal in this basis, as any off-diagonal element only contributes positively to the sum. 

Thus our problem is reduced to solving for the eigenvectors and eigenvalues of $\mu$ and picking $d$ non-negative eigenvalues for $\rho$ that minimize the sum in \eqnref{eq: logL}. Let the eigenvalues of $\mu$ ($\rho$) be $\mu_{i}$ ($\rho_{i}$) and arrange them such that $\mu_{i} \geq \mu_{i+1}$. We then want to minimize $\sum_{i} (\rho_{i} - \mu_{i})^2$ subject to the constraints: $\sum_{i} \mu_{i} = \sum_{i} \rho_{i} = 1$ (which can be imposed through a Lagrange multiplier) and $\rho_{i} \geq 0$ (which is enforced by writing $\rho_{i} = x_{i}^{2}$. This gives the objective function
\begin{equation}
\Lambda = \sum_{i} (x_{i}^{2} - \mu_{i})^{2} - \lambda (\sum_{i} x_{i}^{2} - 1),
\end{equation}
to be optimized with respect to $x_{i}$ and $\lambda$. Differentiating with respect to $x_{i}$, we get
\begin{equation}
\frac{\partial \Lambda}{\partial x_{i}} = 4(x_{i}^2 - \mu_{i})x_{i} -  2\lambda x_{i} = 0,
\end{equation}
which implies that either $x_{i} = 0$ or 
\begin{equation} \label{eq: sol}
x_{i}^2 = \frac{\lambda}{2}  + \mu_{i}
\end{equation}
To find $\lambda$, we pick a set $S = \lbrace i | x_{i} \neq 0 \rbrace$. Summing \eqnref{eq: sol} over $i$ gives
\begin{equation}
\frac{\lambda}{2} = \frac{1}{|S|}\sum_{i \not\in S} \mu_{i}
\end{equation}
and
\begin{equation}
\Lambda(S) = \frac{1}{|S|} \left( \sum_{i \not\in S}\mu_{i} \right)^{2} + \sum_{i \not\in S}\mu_{i}^2
\end{equation}
The task now is to pick the set $S$ which minimises the objective function $\Lambda$ and which satisfies the nonnegativity condition $\lambda/2  + \mu_{i} > 0 $.

One can show that if $\mu_{i} > \mu_{j}$ and $i \in S$ then $\Lambda(S) \leq \Lambda(S + \lbrace j \rbrace - \lbrace i \rbrace)$.  Additionally, we can also show that removing any element from the set $S$ necessarily increases $\Lambda$. Since we have arranged the eigenvalues $\mu_{i}$ in increasing order, the previous two statements tell us that the only task left is to pick the largest number $k$ where $S = \lbrace i | i<k \rbrace$ such that the non-negativity condition remains satisfied. For more details on the proofs for the various statements and the computational complexity of each step refer to the original paper \cite{SmolinGambetta2012}.

\subsection{Entanglement entropies}
\label{supp: entanglement}

The von Neumann entropy $S_{1,A}$ and the second Renyi entropy $S_{2,A}$ for trivial paramagnet, cat state, and topological paramagnet, obtained from the measured density matrices are given by
\begin{center}
\begin{tabular}{|l |c |c |c |c |}
\hline
& exact $S_{1,A}$&measured $S_{1,A}$&exact $S_{2,A}$&measured $S_{2,A}$\\
\hline\hline
$\ket{\mathrm{PM}}$&0&0.41(1)&0&0.19(1)\\
\hline
$\ket{\mathrm{Cat}}$&$\log\,2\approx 0.69$&1.65(2)&$\log\,2\approx 0.69$&1.36(2)\\
\hline
$\ket{\mathrm{SPT}}$&$2\log\,2\approx 1.39$&2.08(2)&$2\log\,2\approx 1.39$&1.92(2)\\
\hline
\end{tabular}
\end{center}
Here, region $A$ consists of 4 adjacent sites of the 8-site chain.
Errors are obtained by repeating the computation with density matrices drawn from their statistical distribution that results from measurement noise and errors. All measured entropies are larger than the exact theoretical values. We attribute this mainly to the decoherence of the quantum state, which in the infinite time limit would lead to a fully mixed state. Also, the readout and gate errors are not accounted for and contribute to deviations of the entanglement entropies from the exact values.

\end{widetext}
\end{document}